\documentclass[%
preprint,
nofootinbib,
amsmath,amssymb,
aps,
pra,
]{revtex4-2}

\usepackage{color}
\usepackage{graphicx}
\usepackage{dcolumn}
\usepackage{bm}

\def\bra#1{\left\langle\,#1\,\right|}
\def\ket#1{\left|\,#1\,\right\rangle}

\renewcommand{\theequation}{\arabic{section}.\arabic{equation}}

\usepackage{tikz}
\newcommand{\mnormalordering}{\hskip 1pt\tikz[baseline]{\node[draw,circle,inner sep=1pt] at (0,0.04) {}; \node[draw,circle,inner sep=1pt] at (0,0.2) {};}\hskip 1pt}

\begin{document}

\title{Defining Root-$T\overline{T}$ operator}

\author{Leszek Hadasz}%
\email{leszek.hadasz@uj.edu.pl, ORCID:0000-0002-8142-8185}
\affiliation{
Institute of Theoretical Physics,
Jagiellonian University,
ul.\ prof.\ Stanis\l{}awa\ \L{}ojasiewicza 11,
30-348~Krak\'ow, Poland}

\author{Rikard von Unge}%
\email{unge@physics.muni.cz, ORCID:0000-0002-8430-2473 }
\affiliation{
Department of Theoretical Physics and Astrophysics\\
Faculty of Science, Masaryk University\\
Kotl\'a\v{r}sk\'a 2, CS-61137 Brno, Czechia
\\
\\
}


\begin{abstract}
We give a tentative definition of the recently introduced Root-$T\bar{T}$ operator in a generic, two dimensional {\em quantum} conformal field theory with continuous spectrum of scaling weights. The definition assumes certain factorization properties and uses Schwinger parametrization to introduce the square root. Properties of the operator thus defined are investigated by explicit computation of variations of two- and three-point correlation functions.
\end{abstract}


\begin{titlepage}
\maketitle
\end{titlepage}

\section{Introduction}
\label{sec:introduction}
Already some time ago \cite{Zamolodchikov:2004ce} it was realized that the so called $T\overline{T}$ operator (or more precisely $\det(T_{\mu\nu})$), constructed from components of the energy momentum tensor in any two dimensional quantum field theory, has very special properties. This operator was more recently used \cite{Smirnov:2016lqw,Cavaglia:2016oda} as an irrelevant perturbation giving rise to families of quantum field theories with special properties following from the properties of the $T\overline{T}$ operator. For instance, the finite volume spectrum of the perturbed theory can be exactly determined in terms of the spectrum of the unperturbed theory. Also, the perturbation preserves integrability \cite{Smirnov:2016lqw,Hernandez-Chifflet:2019sua}. This approach can be further generalized by utilizing a conserved current $(J,\bar{J})$ in addition to the energy momentum tensor \cite{Guica:2017lia}. It is also interesting to observe that the $T\overline{T}$ deformed theory of free fermions is the Volkov-Akulov model with supersymmetry nonlinearly realized \cite{Cribiori:2019xzp}.

A further, recent development inspired by the $T\overline{T}$ program is the definition and study of the so called Root-$T\overline{T}$ operator \cite{Rodriguez:2021tcz, Babaei-Aghbolagh:2022uij, Ferko:2022cix} (see also \cite{Conti:2022egv,Babaei-Aghbolagh:2022leo,Tempo:2022ndz}). It is defined as
\begin{align}
{\mathcal R} = \int\hskip -3pt d^2z\, R(z,\bar{z}) \;,
\end{align}
where
\begin{align}
R(z,\bar{z}) = \sqrt{-\det{\tilde{T}_{\mu\nu}}}
\end{align}
and $\tilde{T}_{\mu\nu} = T_{\mu\nu} - \frac{1}{2}g_{\mu\nu}T$ is the trace-less energy momentum tensor. Due to the presence of the square root in the definition, the operator $R(z,\bar{z})$ has mass dimension $2$ and $\int\! d^2 z\, {\cal R}$ is a classically marginal operator, giving the interesting possibility of constructing families of conformal field theories by perturbing known theories with the Root-$T\overline{T}$ operator. In \cite{Babaei-Aghbolagh:2022uij,Ferko:2022cix} the Lagrangian for Root-$T\overline{T}$ deformed theories of free scalar and fermion fields were found and it was shown that they show similarities with the so-called ModMax theories \cite{Bandos:2020jsw,Babaei-Aghbolagh:2022leo,Garcia:2022wad} of electrodynamics in four dimensions. In \cite{Borsato:2022tmu} it was furthermore shown that the Root-$T\overline{T}$ deformation preserve classical integrability in several interesting cases.

A major difference between the $T\overline{T}$ operator of \cite{Zamolodchikov:2004ce} and the Root-$T\overline{T}$ operator of \cite{Ferko:2022cix} is that the $T\overline{T}$ operator is well defined even quantum mechanically. It can be shown to be free of short distance singularities using point splitting or some other method of regularization. In the case of the Root-$T\overline{T}$ operator there are essentially two obstacles to defining the operator quantum mechanically. Firstly, the presence of the square root makes the definition of the operator into a much more demanding task, and secondly one has to deal with the issue of the short distance singularities when moving products of $T$ and $\overline{T}$ operators together. Some initial steps towards a quantum mechanical definition of the Root-$T\overline{T}$ operator were taken in \cite{Ebert:2023tih} where the problem was discussed in a holographical setting (see also \cite{Tian:2024vln} for a more recent attempt). Even though the question of the precise quantum definition of the operator was not discussed, some conjectures of the quantum behavior of the perturbed theory were given. In particular, by investigating the density of states at large energy and high temperature and comparing to the Cardy formula \cite{Cardy:1986ie,Hartman:2014oaa,Pal:2019zzr} which gives the asymptotic scaling of states in any CFT, the authors conjectured that the high energy limit of the perturbed theory still has the Cardy behavior of a conformal theory although the central charge is changed $c_{\mathrm{eff}} = c e^{-\mu}$ where $\mu$ is the flow parameter. This would indicate that the central charge decreases along the Root-$T\overline{T}$ flow and thus that the quantum Root-$T\overline{T}$ operator would be marginally relevant. 
Different attempts to clarify how to treat the Root-$T\overline{T}$ operator in the quantum theory have been put forward. Recently, the Root-$T\overline{T}$ operator has been investigated for the quantum theory of Root-$T\overline{T}$ deformed chiral bosons in \cite{Ebert:2024zwv}. In \cite{Bagchi:2024unl} the authors used the fact that in CFT's of the Sugawara type the Root-$T\overline{T}$ operator can be written as a current-current bi-linear operator and in \cite{Babaei-Aghbolagh:2024hti,Tsolakidis:2024wut} the Root-$T\overline{T}$ together with the $T\overline{T}$ operator is introduced through a coupling to massive gravity.

Given the recent advances in the field of Root-$T\overline{T}$ deformations and that the classical deformations connect several interesting classes of theories, we feel that it
is of utmost importance to investigate the question of the proper quantum mechanical definition
of the deforming operator. In particular, if a well defined marginal operator can be found, it could be used to find new families of Conformal Field Theories. In this paper we take a first step towards this goal and, although we do not resolve the question, our findings clearly point out where the problem is and indicate a way forward. The answer to this question is essential to the future development of this field. 

We start from a CFT with the action $S[\phi],$ which we perturb by adding a term $\epsilon\int\! d^2z\, R(z,\bar{z})$  with  $\epsilon$ being a dimensionless, infinitesimal parameter. In the path integral this leads to the correlation functions being defined as
\begin{eqnarray}
&&
\hskip -2cm
\left\langle
V_m(z_m,\bar{z}_m) \ldots 
V_1(z_1,\bar{z}_1)
\right\rangle_\epsilon =
\int\hskip -2pt D\phi\; \mathrm{e}^{iS + \epsilon\int R}
\;\;
V_m(z_m,\bar{z}_m)\ldots
V_1(z_1,\bar{z}_1) \\
&=&
\int\hskip -2pt D\phi\; \mathrm{e}^{iS}
\;\;(1+\epsilon \int R + \ldots)
V_m(z_m,\bar{z}_m)\ldots
V_1(z_1,\bar{z}_1) \nonumber \\
& = &
\left\langle
V_m(z_m,\bar{z}_m)\ldots
V_1(z_1,\bar{z}_1)
\right\rangle_0 
+
\epsilon
\int\hskip -2pt d^2z\, \left\langle R(z,\bar{z})
V_m(z_m,\bar{z}_m)\ldots
V_1(z_1,\bar{z}_1)
\right\rangle_0,
\nonumber
\end{eqnarray}
showing that
\begin{equation}
\delta_\epsilon \left\langle V_m(z_m,\bar{z}_m)\ldots
V_1(z_1,\bar{z}_1)
\right\rangle
=
\epsilon
\int\hskip -2pt d^2z\, \left\langle R(z,\bar{z})
V_m(z_m,\bar{z}_m)\ldots
V_1(z_1,\bar{z}_1)
\right\rangle_0,
\end{equation}
where the correlation function on the right hand side is computed in the unperturbed CFT.

To calculate the correlation function we need to somehow implement the problematic square root present in the definition of \(\mathcal{R}\). This is done by using Schwinger parametrization. Essentially we define a power of a generic operator $\mathcal{O}$ as
\begin{equation}
({\mathcal O})^{\alpha} = \frac{1}{\Gamma(-\alpha)}\int\limits_0^\infty \hskip -3pt ds\, s^{-\alpha-1} \mathrm{e}^{-s{\mathcal O}}
=
\frac{1}{2\Gamma(-\alpha)\left(\mathrm{e}^{2\pi i\alpha}-1\right)}\oint\limits_{\mathbb{R}_+}\hskip -2pt ds\, s^{-\alpha-1} \mathrm{e}^{-s{\mathcal O}},
\end{equation}
where the first formula is valid for $\Re\alpha<0$ and the second, with the integration contour encircling the positive real semi-axis in the positive (counterclockwise) direction, gives its analytic continuation
to $s\in\mathbb{C}\!\setminus\!{\mathbb Z}_{\geqslant 0}.$ In particular
\begin{equation}
({\mathcal O})^{\frac12} 
=
\frac{1}{4\sqrt{\pi}}\oint\limits_{\mathbb{R}_+}\hskip -2pt ds\, s^{-\frac32} \mathrm{e}^{-s{\mathcal O}}.
\end{equation}

We also have to make a choice in how to write the operator $\sqrt{T\overline{T}}$. Classically
\begin{equation}
\label{eq:rootTbarTseparation}
\sqrt{T\overline{T}} = \sqrt{\vphantom{\overline{T}}T}\sqrt{\overline{T}},
\end{equation}
and we may choose either as our starting point. It turns out to be computationally advantageous to separate the holomorphic sector from the anti-holomorphic one. 
Such a separation holds for a CFT on a complex plane and in this situation we define
\begin{equation}
({ T\overline{T}})^{\alpha} \equiv \frac{1}{\Gamma(-\alpha)}\int_0^\infty\hskip -5pt ds\, s^{-\alpha-1} :\!\mathrm{e}^{-s{T}}\!: \; \times \;\;
\frac{1}{\Gamma(-\alpha)}\int_0^\infty\hskip -5pt d\bar{s}\, \bar{s}^{-\alpha-1} :\!\mathrm{e}^{-\bar{s}\overline{T}}\!:,
\end{equation}
where we have also normal ordered all products of operators to make them well defined. 

Out of the possible normal ordering prescriptions we use the standard one \cite{diFrancesco}, namely for
\begin{equation}
T(z) = T_-(z) + T_+(z), \hskip 1cm T_-(z)= \sum\limits_{n=-\infty}^{-2}L_nz^{-n-2}
\end{equation}
we define 
\begin{eqnarray}
\label{eq:normal_ordering_two_T}
\nonumber
:\!T(z)T(w)\!:\; & = & T_-(z)T_-(w) + T_-(z)T_+(w) + T_-(w)T_+(z) + T_+(z)T_+(w)
\\
& = &
T(z)T(w) -[T_+(z),T_-(w)]
\end{eqnarray}
and
\begin{equation}
\label{eq:normal_ordering_multiple_T}
:\!T^n(z)\!: = \sum\limits_{k=0}^n {n \choose k}T_-^{n-k}(z)T^{k}_+(z).
\end{equation}
Since
\begin{eqnarray}
\nonumber
\label{eq:Tplus_Tminus_commutator}
[T_+(z),T_-(w)]
& = &
\frac{c}{2(z-w)^4}
+
\frac{2}{(z-w)^2}\left(T_-(w)+T_+(z)\right)
+
\frac{1}{z-w}\left(\partial_w T_-(w) -\partial_z T_+(z)\right)
\\
& = &
\frac{c}{2(z-w)^4}
+
\left(\frac{2}{(z-w)^2} + \frac{1}{z-w}\frac{\partial}{\partial w}\right)T(w)
+ \mathcal{O}(z-w),
\end{eqnarray}
the normal ordering prescription we are using is equivalent to subtracting in the $T(z)T(w)$ (resp.\ $\overline{T}(\bar z)\overline{T}(\bar w)$) operator product expansion terms singular in the $z\to w$ limit.

Our paper is organized as follows. Section \ref{sec:general_case} starts with a general discussion of the proposed definition of the Root-$T\overline{T}$ deformation for a correlation function of arbitrary number of primary fields. 
Then, in subsection  \ref{ssec:two_point_function}, we present an analytic formula for the variation of a two-point correlation function with the primary field $V_\Delta$ of conformal dimension $\Delta$ and derive the form of the change of $\Delta.$
Subsection \ref{ssec:three_point_function} is devoted to the discussion of the form of the variation of the three-point correlation function obtained within the proposed method.  
In section \ref{sec:conclusions} we summarize our results and discuss a possible modification of the applied method of defining the Root-$T\overline{T}$ deformation together with some open problems.

We close the paper with three appendices. In Appendix \ref{ssec:semi_classical_limit} we discuss semi-classical (simultaneously large scaling dimensions) expansion of the obtained formulas, giving an argument in support of the decomposition $\sqrt{T\overline{T}} = \sqrt{T}\sqrt{\overline{T}}.$ We also re-derive here (by summing up perturbative contributions) a formula for the change of the primary field's scaling dimension $\Delta$ under the Root-$T\overline{T}$ deformation. 
Appendix \ref{ssec:free_scalar} contains a discussion of the semi-classical limit of the Root-$T\overline{T}$ variation of the free, massless boson CFT and serves as a consistency check of  our constructions.
Finally, Appendix \ref{app:analytic_properties} contains proofs of the properties of the variation of the two-point function used in subsection \ref{ssec:two_point_function}.

\section{The general case}
\setcounter{equation}{0}
\label{sec:general_case}
Let $T(z)$ be the holomorphic component of the stress tensor of a given two dimensional conformal field theory. Following the discussion in the introduction, we would like to calculate the $z$ dependence of correlation functions of the type
\begin{equation}
\bra{0}:\hskip -2pt\mathrm{e}^{-sT(z)}\hskip -2pt:V_{\Delta_m}(w_m,\bar w_m)\ldots V_{\Delta_1}(w_1,\bar w_1)\ket{0},
\end{equation}
with $\ket{0}$ being the SL(2,$\mathbb{C}$) invariant vacuum and $V_{\Delta_j}(w_j,\bar w_j)$ a CFT primary field with the scaling dimension $\Delta_j\in \mathbb{R}$ (we assume in the present paper that the holomorphic and antiholomorphic scaling dimensions of the field $V$ are equal and real).
The $z$-dependence of the correlation function
\begin{equation}
\label{eq:general_correlation_function}
\bra{0}T(z)V_{\Delta_m}(w_m,\bar w_m)\ldots V_{\Delta_1}(w_1,\bar w_1)\ket{0}
\end{equation}
is fully determined by the conformal Ward identity
\begin{eqnarray}
\label{eq:CWI:1}
&&
\hskip -2cm
\bra{0}T(z)V_{\Delta_m}(w_m,\bar w_m)\ldots V_{\Delta_1}(w_1,\bar w_1)\ket{0}
\\[4pt]
\nonumber
& = &
\sum\limits_{j=1}^m\left(-\frac{1}{w_j-z}\frac{\partial}{\partial w_j} + \frac{\Delta_j}{(w_j-z)^2}\right)
\bra{0}V_{\Delta_m}(w_m,\bar w_m)\ldots V_{\Delta_1}(w_1,\bar w_1)\ket{0},
\end{eqnarray}
so that the action of $T(z)$ can be represented as a differential operator.
We adopt the convention that the principal value of the argument of a complex number takes value in the interval $[0,2\pi)$ and assume that $\Re(w -z) > 0,$ so that
\begin{eqnarray}
\label{eq:derivative_simple_form}
\nonumber
&&
\hskip -2cm
\left(-\frac{1}{w-z}\frac{\partial}{\partial w} + \frac{\Delta}{(w-z)^2}\right)f(w)
=
(w-z)^\Delta \left(-\frac{1}{w-z}\frac{\partial}{\partial w}\right)\left((w-z)^{-\Delta}f(w)\right)
\\[4pt]
& = &
(2t)^{\frac{\Delta}{2}}\left.\left(-\frac{\partial}{\partial t}\right)\left((2t)^{-\frac{\Delta}{2}}f\left(z+\sqrt{2t}\right)\right)\right|_{t = \frac12(w-z)^2}.
\end{eqnarray}
This differential operator can now be easily exponentiated
\begin{eqnarray}
\label{eq:derivative_exponential}
\nonumber
&&
\hskip -1cm
\exp \left\{-s\left(-\frac{1}{w-z}\frac{\partial}{\partial w} + \frac{\Delta}{(w-z)^2}\right)\right\}f(w)
=
\sum\limits_{n=0}^\infty \frac{(-s)^n}{n!}\left(-\frac{1}{w-z}\frac{\partial}{\partial w} + \frac{\Delta}{(w-z)^2}\right)^n f(w)
\\[4pt]
\nonumber
& = &
(2t)^{\frac{\Delta}{2}}\left.\exp\left\{s\frac{\partial}{\partial t}\right\}\left((2t)^{-\frac{\Delta}{2}}f\left(z+\sqrt{2t}\right)\right)\right|_{t = \frac12(w-z)^2}
\\[4pt]
& = &
(2t)^{\frac{\Delta}{2}}\left.(2t+2s)^{-\frac{\Delta}{2}}f\left(z+\sqrt{2t+2s}\right)\right|_{t = \frac12(w-z)^2}
\\[4pt]
\nonumber
& = &
(w-z)^\Delta \left((w-z)^2+2s\right)^{-\frac{\Delta}{2}}f\left(z+\sqrt{(w-z)^2+2s}\right).
\end{eqnarray}
As discussed in the introduction, we normal order our operator $:\!\mathrm{e}^{-sT(z)}\!\!:$ by subtracting singularities that are present in the products of stress tensors evaluated at the same point or, equivalently, by standard normal ordering of the modes.
Then, using (\ref{eq:derivative_exponential}), we have
\begin{eqnarray}
\label{eq:exponential:of:T(z)}
\nonumber
&&
\hskip -.5cm
\bra{0}:\!\exp\big\{-sT(z)\big\}\!:V_{\Delta_m}(w_m,\bar w_m)\ldots V_{\Delta_1}(w_1,\bar w_1)\ket{0}
\\[4pt]
\nonumber
& = &
\sum\limits_{n=0}^\infty \frac{(-s)^n}{n!}\lim\limits_{z_i\to z}\bra{0}:\!T(z_n)\ldots T(z_1)\!:V_{\Delta_m}(w_m,\bar w_m)\ldots V_{\Delta_1}(w_1,\bar w_1)\ket{0}
\\
& = &
\exp\left\{-s\sum\limits_{j=1}^m\left(-\frac{1}{w_j-z}\frac{\partial}{\partial w_j} + \frac{\Delta_j}{(w_j-z)^2}\right) \right\}
\bra{0}V_{\Delta_m}(w_m,\bar w_m)\ldots V_{\Delta_1}(w_1,\bar w_1)\ket{0}
\\
\nonumber
& = &
\prod\limits_{j=1}^m(w_j-z)^{\Delta_j} \left((w_j-z)^2+2s\right)^{-\frac{\Delta_j}{2}}
\bra{0}V_{\Delta_m}\left(\xi_m,\bar w_m\right)\ldots V_{\Delta_1}\left(\xi_1,\bar w_m\right)\ket{0}\hskip -2pt\Big|_{\xi_j = z+\sqrt{(w_j-z)^2+2s}}
\end{eqnarray}
and analogously in the anti-holomorphic sector.

We now \textbf{define} the action of the Root-$T\overline{T}$ operator within the correlation functions of primary fields 
as
\begin{eqnarray}
\label{eq:root:TbarT:definition:1}
&&
\hskip -.5cm
\int\limits_{\mathrm{\scriptscriptstyle reg.}}\!d^2z
\bra{0}\!:\!T^\alpha(z)\!: :\!\overline{T}{}^\alpha(\bar z)\!:\hskip -2pt\prod\limits_{j=1}^m V_{\Delta_j}(w_j,\bar w_j)\ket{0}\Big|_{\alpha= \frac12}
\\[4pt]
\nonumber
& = & 
\bra{0}\hskip -2pt\prod\limits_{j=1}^m\hskip -1pt V_{\Delta_j}(w_j,\bar w_j)\ket{0}\hskip -3pt
\int\limits_{\mathrm{\scriptscriptstyle reg.}}\hskip -3pt d^2z\,
\Bigg|
\frac{1}{\bra{0}\prod\limits_{j=1}^m V_{\Delta_j}(w_j,\bar w_j)\ket{0}}\bra{0}\hskip -3pt :\!T^\alpha(z)\!:\hskip -2pt\prod\limits_{j=1}^m V_{\Delta_j}(w_j,\bar w_j)\hskip -2pt\ket{0}\Big|_{\alpha= \frac12}
\Bigg|^2
\end{eqnarray}
with
\begin{eqnarray}
\label{eq:root:TbarT:definition:2}
&&
\hskip -2cm 
\frac{1}{\bra{0}\prod\limits_{j=1}^m V_{\Delta_j}(w_j,\bar w_j)\ket{0}}\ 
\bra{0}:\!T^\alpha(z)\!:\prod\limits_{j=1}^m V_{\Delta_j}(w_j,\bar w_j)\ket{0}\Big|_{\alpha= \frac12}
\\
\nonumber
& = & 
\frac{1}{\bra{0}\prod\limits_{j=1}^m V_{\Delta_j}(w_j,\bar w_j)\ket{0}}\ 
\oint\limits_{\mathbb{R}_+}\! \frac{ds}{4\sqrt{\pi}} s^{-\frac32}
\bra{0}:\!\mathrm{e}^{-sT(z)}\!:\prod\limits_{j=1}^m V_{\Delta_j}(w_j,\bar w_j)\ket{0}.
\end{eqnarray}
Since in a given CFT on the complex plane the correlation functions of primary fields are known (explicitly for $m=2$ and $m=3$ and in terms of sums/integrals containing conformal blocks  for $m \geqslant 4$), formulas (\ref{eq:exponential:of:T(z)})-- (\ref{eq:root:TbarT:definition:2}) allows us to compute the variation of the correlation functions in question under the Root-$T\overline{T}$ deformation.

\subsection{The two-point function}
\label{ssec:two_point_function}
By global conformal invariance, the two-point correlation function of the primary fields $V_{\Delta_2}(w_2,\bar w_2)$ and $V_{\Delta_1}(w_1,\bar w_1)$ vanishes unless
their scaling dimensions are equal. When this condition is met we have
\begin{equation}
\label{eq:two:point:correlation:function}
\bra{0}V_{\Delta}(w_2,\bar w_2)V_{\Delta}(w_1,\bar w_1)\ket{0} = \frac{\mathcal{N}_\Delta^2}{(w_2-w_1)^{2\Delta}(\bar w_2 - \bar w_1)^{2\Delta}},
\end{equation}
where $\mathcal{N}_\Delta$ is a normalization constant, not fixed by the conformal invariance. For the  correlation function of this form the formula (\ref{eq:exponential:of:T(z)}) gives
\begin{eqnarray}
\label{eq:exponential:of:T:2pt}
&&
\hskip -1.5cm
\frac{\bra{0}:\!\mathrm{e}^{-sT(z)}\!:V_{\Delta}(w_2,\overline{w}_2)V_\Delta(w_1,\overline{w}_1)\ket{0}}
{\bra{0}V_{\Delta}(w_2,\overline{w}_2)V_\Delta(w_1,\overline{w}_1)\ket{0}}
\\[4pt]
\nonumber
& = &
\prod\limits_{j=1}^2\left(\frac{w_j-z}{\sqrt{(w_j-z)^2+2s}}\right)^{\Delta}
\left(\frac{w_2-w_1}{\sqrt{(w_2-z)^2+2s}-\sqrt{(w_1-z)^2+2s}}\right)^{2\Delta}
\\[4pt]
\nonumber
& = &
\left(\sqrt{\frac{x_1}{x_2}} + \sqrt{\frac{x_2}{x_1}}\right)^{-2\Delta}
\left(\sqrt[4]{\frac{x_2^2+2s}{x_1^2+2s}} + \sqrt[4]{\frac{x_1^2+2s}{x_2^2+2s}}\right)^{2\Delta},
\end{eqnarray}
where $x_j = w_j-z,\ j = 1,2.$ The possible phase ambiguities are fixed by the assumption that $\Re\,w_2 > \Re\,w_1,$ and we define the r.h.s.\ of (\ref{eq:exponential:of:T:2pt}) as a meromorphic function of $z$ by analytic continuation from the region $\Re\,w_1 > \Re\,z.$ 

In the defining formula  (\ref{eq:root:TbarT:definition:2}) we integrate around the cut of the $s^{-\frac32}$ function, which we choose to be the semi-axis of real, positive $s.$ On the other hand, the function
\begin{equation}
\sqrt[4]{\frac{x_2^2+2s}{x_1^2+2s}} + \sqrt[4]{\frac{x_1^2+2s}{x_2^2+2s}}  
\end{equation}
has branch points at $2s= -x_i^2$ and can be defined as an analytic function of $s \in \mathbb{C}\setminus \{[-x_2^2,-x_1^2]\}.$ We may deform the contour around the cut on the positive real axis to a contour around the new cut. 
In the vicinity of this cut we have
\begin{equation}
\sqrt[4]{\frac{x_1^2 + 2s}{x_2^2+2s}} + \sqrt[4]{\frac{x_2^2 + 2s}{x_1^2+2s}}
\; \to \;
\mathrm{e}^{\pm\frac{i\pi}{4}}\sqrt[4]{\frac{|x_1^2 +2s|}{|x_2^2+2s|}} + \mathrm{e}^{\mp\frac{i\pi}{4}}\sqrt[4]{\frac{|x_2^2+2s|}{|x_1^2+2s|}},
\end{equation}
where the upper/lower sign corresponds to $s$ approaching the cut from above/below.
Denoting
\begin{equation}
\label{eq:p:parameter}
p = \sqrt[4]{\frac{|x_1^2+2s|}{|x_2^2+2s|}},
\end{equation}
and choosing $\theta$ such that:
\begin{equation}
\cos\frac12\theta = \frac{p+p^{-1}}{\sqrt{2\left(p^2+p^{-2}\right)}}, \hskip 1cm \sin\frac12\theta = \frac{p-p^{-1}}{\sqrt{2\left(p^2+p^{-2}\right)}},
\end{equation}
we have
\begin{equation}
\left(\mathrm{e}^{\pm\frac{i\pi}{4}}\sqrt[4]{\frac{|x_1^2+2s|}{|x_2^2+2s|}} + \mathrm{e}^{\mp\frac{i\pi}{4}}\sqrt[4]{\frac{|x_2^2+2s|}{|x_1^2+2s|}}\right)^{2\Delta}  
\hskip -5pt
=
\left(\sqrt{\frac{|x_1^2+2s|}{|x_2^2+2s|}} + \sqrt{\frac{|x_2^2+2s|}{|x_1^2+2s|}}\right)^{\Delta}
\hskip -5pt
\left(\mathrm{e}^{\pm i\theta}\right)^\Delta.
\end{equation}

\vskip 3mm
\begin{figure}[htb]
\caption{Deforming the contour from one cut to the other through infinity}
\vskip 3mm
\centerline{\includegraphics[width=.8\textwidth]{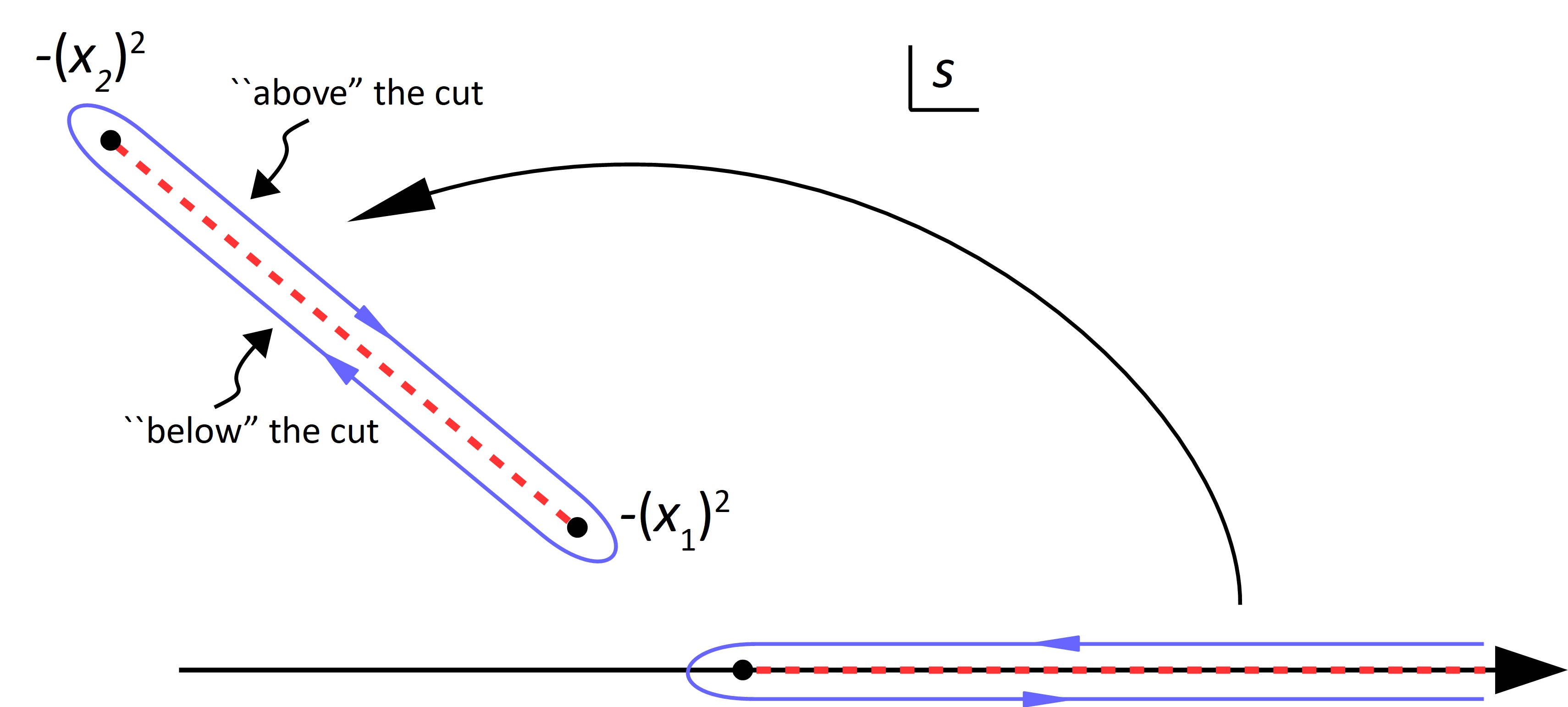}}
\label{fig:cut}
\end{figure}
\vskip 3mm

Substituting $s = \mathrm{e}^{i\pi}t$ and deforming the contour onto the cut as illustrated in figure \ref{fig:cut} (the contribution from the circle at infinity vanishes) we thus have\footnote{
As it stands, the integral representation (\ref{eq:sqrtT:cut:integral}) is valid for $\Delta < 2.$ If $\Delta \geq 2,$ then the deformed contour is no longer a line segment and the divergent contributions from $t \to \frac12 x_j^2$ are subtracted by contributions from the integrals along small circles centered at these points.}
\begin{eqnarray}
\label{eq:sqrtT:cut:integral}
&&
\hskip -1cm
\frac{\bra{0}\sqrt{T(z)}V_\Delta(w_2,\bar w_2)V_{\Delta}(w_1,\bar w_1)\ket{0}}{\bra{0}V_\Delta(w_2,\bar w_2)V_{\Delta}(w_1,\bar w_1)\ket{0}}
\\[4pt]
\nonumber
& = &
\frac{1}{2\sqrt{\pi}}\left(\sqrt{\frac{x_2}{x_1}} + \sqrt{\frac{x_1}{x_2}}\right)^{-2\Delta}\int\limits_{\frac12x_1^2}^{\frac12x_2^2}\hskip -3pt \frac{dt}{\sqrt{t^3}}
\left(\sqrt{\left|\frac{2t-x_1^2}{x_2^2-2t}\right|} + \sqrt{\left|\frac{x_2^2 -2t}{2t-x_1^2}\right|}\right)^{\Delta}\sin\Big(\Delta\;\theta(x_1,x_2)\Big).
\end{eqnarray}
As an explicit example supporting our general conclusions let us take the simplest case of $\Delta = 1$ and real, positive $x_1^2 < x_2^2.$  The r.h.s.\ of (\ref{eq:sqrtT:cut:integral}) can now be presented as
\begin{eqnarray}
&&
\hskip -2cm
\frac{1}{2\sqrt{\pi}}\left(\sqrt{\frac{x_2}{x_1}} + \sqrt{\frac{x_1}{x_2}}\right)^{-2}\int\limits_{\frac12x_1^2}^{\frac12x_2^2}\hskip -3pt dt\ t^{-\frac32}
\left(\sqrt{\frac{x_2^2 -2t}{2t-x_1^2}} - \sqrt{\frac{2t-x_1^2}{x_2^2-2t}}\right)
\\[4pt]
& = &
\nonumber
\sqrt{\frac{2}{\pi}}\frac{x_2}{(x_1+x_2)^2}
\left(\left(1+\frac{x_1^2}{x_2^2}\right)E\left(1-\frac{x_2^2}{x_1^2}\right) - 2K\left(1-\frac{x_2^2}{x_1^2}\right)\right),
\end{eqnarray}
where
\begin{equation}
E(x) = \int\limits_0^{\frac{\pi}{2}}\!d\theta\ \sqrt{1-x\sin^2\theta},
\hskip 1cm
K(x) = \int\limits_0^{\frac{\pi}{2}} \frac{d\theta}{\sqrt{1-x\sin^2\theta}},
\end{equation}
are complete elliptic integrals. Notice that for $y\to\infty:$
\begin{equation}
E(1-y) = \sqrt{y} + \frac14\frac{\log y}{\sqrt{y}} + \mathcal{O}\left(y^{-\frac12}\right),
\hskip 1cm
K(1-y) = \frac12\frac{\log y}{\sqrt{y}} + \mathcal{O}\left(y^{-\frac12}\right),
\end{equation}
so that, for $z\to w_1,$
\begin{equation}
\frac{\bra{0}\sqrt{T(z)}V_1(w_2,\bar w_2)V_1(w_1,\bar w_1)\ket{0}}{\bra{0}V_1(w_2,\bar w_2)V_1(w_1,\bar w_1)\ket{0}}
=
\sqrt{\frac{2}{\pi}}\frac{1}{w_1-z} +\mathcal{O}\left(\frac{w_1-z}{w_2-z}\log(w_1-z)\right).
\end{equation}
Moreover, since for $y\to 0:$
\begin{equation}
K(y) = \frac{\pi}{2} + \frac{\pi y}{8} + \mathcal{O}\left(y^2\right), \hskip 1cm E(y) = \frac{\pi}{2} - \frac{\pi y}{8} + \mathcal{O}\left(y^2\right),
\end{equation}
we get that for $z \to -\infty$
\begin{equation}
\frac{\bra{0}\sqrt{T(z)}V_1(w_2,\bar w_2)V_1(w_1,\bar w_1)\ket{0}}{\bra{0}V_1(w_2,\bar w_2)V_1(w_1,\bar w_1)\ket{0}}
=
o\left(z^{-2}\right).
\end{equation}

Let us now define
\begin{equation}
\xi = \frac{w_1-z}{w_2-w_1} \hskip 5mm \Rightarrow \hskip 5mm x_1 = w_{21}\xi, \hskip 5mm x_2 = w_{21}(1+\xi),
\end{equation}
and change in (\ref{eq:sqrtT:cut:integral}) the integration variable from $t$ to $u = 2t/w_{12}^2.$ Introducing the short hand notation
\begin{eqnarray}
\label{eq:SHCFdef}
\nonumber
\left\langle \sqrt{T(z)}\right\rangle_{w_2,w_1}
& \equiv &
\frac{\bra{0}\sqrt{T(z)}V_\Delta(w_2,\bar w_2)V_{\Delta}(w_1,\bar w_1)\ket{0}}{\bra{0}V_\Delta(w_2,\bar w_2)V_{\Delta}(w_1,\bar w_1)\ket{0}}
\end{eqnarray}
this yields
\begin{eqnarray}
\label{eq:sqrtT:cut:integral:2}
\nonumber
\left\langle \sqrt{T(z)}\right\rangle_{w_2,w_1}
& = &
\sqrt{\frac{2}{\pi}}\frac{1}{w_2-w_1}\left(\sqrt{\frac{1+\xi}{\xi}} + \sqrt{\frac{\xi}{1+\xi}}\right)^{-2\Delta}
\\[4pt]
\nonumber
& \times &
\int\limits_{\xi^2}^{(1+\xi)^2}\hskip -3pt \frac{du}{\sqrt{u^3}}
\left(\sqrt{\left|\frac{u-\xi^2}{(1+\xi)^2-u}\right|} + \sqrt{\left|\frac{(1+\xi)^2-u}{u-\xi^2}\right|}\right)^{\Delta}\sin\Big(\Delta\tilde\theta(\xi|u)\Big),
\end{eqnarray}
where 
\begin{equation}
\cos\frac12\tilde\theta(\xi|u) = \frac{q+q^{-1}}{\sqrt{2\left(q^2+q^{-2}\right)}}, \hskip .2cm \sin\frac12\tilde\theta(\xi|u) = \frac{q-q^{-1}}{\sqrt{2\left(q^2+q^{-2}\right)}},
\hskip .2cm
q \equiv q(\xi|u) = \sqrt[4]{\left|\frac{u-\xi^2}{(1+\xi)^2-u}\right|}.
\end{equation}
In effect, save for the explicit factor $(w_2-w_1)^{-1},$ (\ref{eq:sqrtT:cut:integral:2}) depends on $w_1$ and $w_2$  solely through the combination $\xi = \frac{w_1-z}{w_2-w_1}.$

\pagebreak
The following properties of this function are proven in Appendix \ref{app:analytic_properties}: 
\begin{itemize}
\item 
we have
\begin{equation}
\label{eq:w2_to_infinity}
\lim\limits_{w_2\to\infty} \left\langle \sqrt{T(z)}\right\rangle_{w_2,w_1} = \frac{\sqrt{2}\,\Gamma\left(\frac{\Delta+1}{2}\right)}{\Gamma\left(\frac{\Delta}{2}\right)}\frac{1}{w_1-z};
\end{equation}
\item
for $z \to w_1:$
\begin{equation}
\label{eq:z_to_w1}
(w_1-z) \left\langle \sqrt{T(z)}\right\rangle_{w_2,w_1} = \frac{\sqrt{2}\,\Gamma\left(\frac{\Delta+1}{2}\right)}{\Gamma\left(\frac{\Delta}{2}\right)} + o(1);
\end{equation}
\item 
for $z \to \infty:$
\begin{equation}
\label{eq:z_to_infinity}
\left\langle \sqrt{T(z)}\right\rangle_{w_2,w_1} = o\left(z^{-1}\right).
\end{equation}
\end{itemize}

Repeating the construction above in the regions corresponding to the other relations between $\Re w_1,\ \Re w_2$ and $\Re z$ we conclude, that the only non-integrable singularities of the function
$\left\langle \sqrt{T(z)}\right\rangle_{w_2,w_1}$ are first order poles at $z=w_1, z= w_2,$ whose residues are (up to sign) equal to $\frac{\sqrt{2}\,\Gamma\left(\frac{\Delta+1}{2}\right)}{\Gamma\left(\frac{\Delta}{2}\right)}.$ The integral
\begin{equation}
Z_1(\Delta)
=
\frac12\int\!d^2z \left(\left|\left\langle \sqrt{T(z)}\right\rangle_{w_2,w_1}\right|^2 
- 2\left|\frac{\Gamma\left(\frac{\Delta+1}{2}\right)}{\Gamma\left(\frac{\Delta}{2}\right)}\right|^2\frac{|w_2-w_1|^2}{|w_2-z|^2|w_1-z|^2}\right)
\end{equation}
is thus convergent and, in view of (\ref{eq:sqrtT:cut:integral:2}), independent of $w_1$ and $w_2.$

On the other hand, the integral
\begin{equation}
\int\hskip -3pt \frac{d^2z\, |w_2-w_1|^2}{|w_2-z|^2|w_1-z|^2}
\end{equation}
is divergent and in order to define it we need to use some regularization procedure. For instance, if we use the cut-off regularization,
\begin{equation}
\int\limits_{\mathrm{\scriptscriptstyle reg.}} \hskip-3pt \frac{d^2z\, |w_2-w_1|^2}{|w_2-z|^2|w_1-z|^2}
=
\int\hskip -3pt \frac{d^2z\, |w_2-w_1|^2}{\left(|w_2-z|^2+\ell^2\right)\left(|w_1-z|^2+\ell^2\right)},
\hskip 5mm \ell^2 > 0,
\end{equation}
then we obtain
\begin{eqnarray}
\nonumber
\int\limits_{\mathrm{\scriptscriptstyle reg.}} \hskip-3pt \frac{d^2z\, |w_2-w_1|^2}{|w_2-z|^2|w_1-z|^2}
& = &
\frac{2\pi|w_2-w_1|}{\sqrt{|w_2-w_1|^2 + 4\ell^2}}\log\frac{\sqrt{|w_2-w_1|^2 + 4\ell^2}+|w_2-w_1|}{\sqrt{|w_2-w_1|^2 + 4\ell^2}-|w_2-w_1|}
\\[4pt]
& \stackrel{\ell\to 0}{=} &
4\pi\log\frac{|w_2-w_1|}{\ell} + {\mathcal O}(\ell^2),
\end{eqnarray}
while if we choose the the dimensional regularization
\begin{equation}
\int\limits_{\mathrm{\scriptscriptstyle reg.}} \hskip-3pt \frac{d^2z\, |w_2-w_1|^2}{|w_2-z|^2|w_1-z|^2}
=
\int\hskip -3pt \frac{d^Dz\, |w_2-w_1|^2}{|w_2-z|^2|w_1-z|^2}, \hskip 5mm 2 < D < 4,
\end{equation}
we get
\begin{eqnarray}
\int\limits_{\mathrm{\scriptscriptstyle reg.}} \hskip-3pt \frac{d^2z\, |w_2-w_1|^2}{|w_2-z|^2|w_1-z|^2}
& = &
\pi^{\frac{D}{2}}|w_2-w_1|^{D-2}  \frac{\Gamma\left(2-\frac{D}{2}\right)\Gamma^2\left(\frac{D}{2}-1\right)}{\Gamma(D-2)}
\\[4pt]
\nonumber
& \stackrel{D\to 2}{=} &
\frac{4\pi}{D-2} + 4\pi\log|w_2-w_1| + 2\pi\left(\gamma_{\mathrm E} + \log\pi\right) + \mathcal{O}\left((D-2)\right).
\end{eqnarray}
We thus arrive at the conclusion that the variation of the two point correlation function 
\begin{equation}
\delta_\epsilon \langle\,0\,|\sqrt{T(z)\overline{T}(\bar z)}V_{\Delta}(w_2,\bar w_2)V_{\Delta}(w_1,\bar w_1)|\,0\,\rangle
=
\epsilon|w_2-w_1|^{-4\Delta}\int\limits_{\mathrm{\scriptscriptstyle reg.}} \hskip-3pt d^2z \left|\left\langle \sqrt{T(z)}\right\rangle_{w_2,w_1}\right|^2
\end{equation}
can be written as a difference
\begin{equation}
\frac{(1+\epsilon Z_{\mathrm{reg.}}(\Delta))^2}{|w_2-w_1|^{4\Delta + 4\epsilon \delta\Delta}} - \frac{1}{|w_2-w_1|^{4\Delta}}
=
\epsilon\frac{2Z_{\mathrm{reg.}}(\Delta) - 4\delta\Delta\log|w_2-w_1|}{|w_2-w_1|^2} + \mathcal{O}(\epsilon^2),
\end{equation}
where
\begin{equation}
Z_{\mathrm{reg.}}(\Delta) = Z_1(\Delta) + \hbox{\rm regulator dependent terms}
\end{equation}
is a re-normalization factor of the $V_{\Delta}(w,\bar w)$ field and 
\begin{equation}
\label{eq:scaling:dimension:variation}
\delta\Delta = -2\pi \frac{\Gamma^2\left(\frac{\Delta+1}{2}\right)}{\Gamma^2\left(\frac{\Delta}{2}\right)}
\end{equation}
is the change of the field's scaling dimension. 

\subsection{The three-point function}
\label{ssec:three_point_function}
The three-point function of primary fields is of the form
\begin{equation}
\bra{0}V_{\Delta_3}(w_3,\bar w_3)V_{\Delta_2}(w_2,\bar w_2)V_{\Delta_1}(w_1,\bar w_1)\ket{0}
=
\frac{\bra{\Delta_3}V_{\Delta_2}(1,1)\ket{\Delta_1}}
{|w_3-w_2|^{2\Delta^1_{23}}|w_3-w_1|^{2\Delta^2_{13}}|w_2-w_1|^{2\Delta^3_{12}}}
\end{equation}
and for real  $0< w_1-z < w_2-z < w_3-z$ the formula (\ref{eq:exponential:of:T(z)}) gives
\begin{eqnarray}
\label{eq:3pt:exponential:of:T(z):explicit}
&&
\hskip -.5cm
\frac{\bra{0}:\!\exp\left\{-sT(z)\right\}\!:V_{\Delta_3}(w_3,\bar w_3)V_{\Delta_2}(w_2,\bar w_2)V_{\Delta_1}(w_1,\bar w_1)\ket{0}}
{\bra{0}V_{\Delta_3}(w_3,\bar w_3)V_{\Delta_2}(w_2,\bar w_2)V_{\Delta_1}(w_1,\bar w_1)\ket{0}}
\\[4pt]
\nonumber
& = &
\left(\sqrt{\frac{x_1}{x_2}} + \sqrt{\frac{x_2}{x_1}}\right)^{-\Delta^3_{12}}
\left(\sqrt{\frac{x_1}{x_3}} + \sqrt{\frac{x_3}{x_1}}\right)^{-\Delta^2_{13}}
\left(\sqrt{\frac{x_2}{x_3}} + \sqrt{\frac{x_3}{x_2}}\right)^{-\Delta^1_{23}}
\\[4pt]
\nonumber
& \times &
\left(\sqrt[4]{\frac{x_2^2+2s}{x_1^2+2s}} + \sqrt[4]{\frac{x_1^2+2s}{x_2^2+2s}}\right)^{\hskip -3pt\Delta^3_{12}}\hskip -5pt
\left(\sqrt[4]{\frac{x_3^2+2s}{x_1^2+2s}} + \sqrt[4]{\frac{x_1^2+2s}{x_3^2+2s}}\right)^{\hskip -3pt\Delta^2_{13}}\hskip -5pt
\left(\sqrt[4]{\frac{x_3^2+2s}{x_2^2+2s}} + \sqrt[4]{\frac{x_2^2+2s}{x_3^2+2s}}\right)^{\hskip -3pt\Delta^1_{23}}\hskip -7pt.
\end{eqnarray}
For $0 < s \leq x_1 \ll x_2,x_3$ this function is approximated by
\begin{equation}
\frac{x_1^{\Delta_1}}{\left(x_1^2+2s\right)^{\frac12\Delta_1}}
\end{equation}
and an argument parallel to the one presented for the two-point function above gives
\begin{equation}
\frac{\bra{0}:\!\sqrt{T(z)}\!:V_{\Delta_3}(w_3,\bar w_3)V_{\Delta_2}(w_2,\bar w_2)V_{\Delta_1}(w_1,\bar w_1)\ket{0}}
{\bra{0}V_{\Delta_3}(w_3,\bar w_3)V_{\Delta_2}(w_2,\bar w_2)V_{\Delta_1}(w_1,\bar w_1)\ket{0}}
\;\stackrel{z\to w_1}{\approx}
\frac{\sqrt{2}\,\Gamma\left(\frac{\Delta_1+1}{2}\right)}{\Gamma\left(\frac{\Delta_1}{2}\right)}\frac{1}{w_1-z}.
\end{equation}
Therefore, if we define
\begin{equation}
R_{\hbox{\boldmath ${\scriptstyle\Delta}$}}\!(\hbox{\boldmath $w$}) = 
\sum\limits_{(a,b,c)}\hskip -3pt\left(
\left|\frac{\Gamma\left(\frac{\Delta_a+1}{2}\right)}{\Gamma\left(\frac{\Delta_a}{2}\right)}\right|^2
+
\left|\frac{\Gamma\left(\frac{\Delta_b+1}{2}\right)}{\Gamma\left(\frac{\Delta_b}{2}\right)}\right|^2
-
\left|\frac{\Gamma\left(\frac{\Delta_c+1}{2}\right)}{\Gamma\left(\frac{\Delta_c}{2}\right)}\right|^2
\right)
\frac{|w_a-w_b|^2}{|z-w_a|^2|z-w_b|^2}
\end{equation}
with $(a,b,c) \in \{(1,2,3),(2,3,1),(3,1,2)\},$ then the function
\begin{equation}
\label{eq:3pt:variation:regularized}
\left|\left\langle\!\sqrt{T(z)}\right\rangle_{\hskip -3pt\hbox{\boldmath{$\scriptstyle w$}}}\right|^2_{\mathrm{reg.}}
=
\Bigg|
\frac{\bra{0}:\!\sqrt{T(z)}\!:V_{\Delta_3}(w_3,\bar w_3)V_{\Delta_2}(w_2,\bar w_2)V_{\Delta_1}(w_1,\bar w_1)\ket{0}}
{\bra{0}V_{\Delta_3}(w_3,\bar w_3)V_{\Delta_2}(w_2,\bar w_2)V_{\Delta_1}(w_1,\bar w_1)\ket{0}}
\Bigg|^2
-
R_{\hbox{\boldmath ${\scriptstyle\Delta}$}}\!(\hbox{\boldmath $w$})
\end{equation}
is integrable over the complex $z$ plane. Using the notation from the previous subsection we can now write
\begin{eqnarray}
\label{eq:3point:variation:new}
\nonumber
&&\hskip -3cm
\int\limits_{\mathrm{\scriptscriptstyle reg.}}\hskip -3pt d^2z\, 
\Bigg|
\frac{\bra{0}:\!\sqrt{T(z)}\!:V_{\Delta_3}(w_3,\bar w_3)V_{\Delta_2}(w_2,\bar w_2)V_{\Delta_1}(w_1,\bar w_1)\ket{0}}
{\bra{0}V_{\Delta_3}(w_3,\bar w_3)V_{\Delta_2}(w_2,\bar w_2)V_{\Delta_1}(w_1,\bar w_1)\ket{0}}
\Bigg|^2
\\
& = & 
\int\hskip -3pt d^2z \left|\left\langle\!\sqrt{T(z)}\right\rangle_{\hskip -3pt\hbox{\boldmath{$\scriptstyle w$}}}\right|^2_{\mathrm{reg.}}
+
\int\limits_{\mathrm{\scriptscriptstyle reg.}}\hskip -3pt d^2z\, R_{\hbox{\boldmath ${\scriptstyle\Delta}$}}\!(\hbox{\boldmath $w$})
\\
\nonumber
& = &
\int\hskip -3pt d^2z \left|\left\langle\!\sqrt{T(z)}\right\rangle_{\hskip -3pt\hbox{\boldmath{$\scriptstyle w$}}}\right|^2_{\mathrm{reg.}}
-
\sum\limits_{a=1}^3 Z_1(\Delta_a)
\\
\nonumber
& + &
\sum\limits_{a=1}^3 Z_{\mathrm{\scriptscriptstyle reg.}}(\Delta_a) 
-
2\sum\limits_{(a,b,c)}\big(\delta\Delta_a + \delta\Delta_b-\delta\Delta_c\big)\log|w_a-w_b|.
\end{eqnarray}
Since
\begin{eqnarray}
&&
\hskip -1cm
\sum\limits_{a=1}^3 Z_{\mathrm{\scriptscriptstyle reg.}}(\Delta_a) 
-
2\sum\limits_{(a,b,c)}\big(\delta\Delta_a + \delta\Delta_b-\delta\Delta_c\big)\log|w_a-w_b|
\\
\nonumber
&= &
\prod\limits_{(a,b,c)}|w_a-w_b|^{2\Delta^c_{ab}}\lim\limits_{\epsilon\to 0}
\frac{1}{\epsilon}
\Bigg(
\frac{\prod_{a=1}^3(1+\epsilon Z_{\mathrm{\scriptscriptstyle reg.}}(\Delta_a))}
{\prod\limits_{(a,b,c)}|w_a-w_b|^{2\Delta^c_{ab}+ 2\epsilon\delta\Delta^c_{ab}}}
-
\frac{1}{\prod\limits_{(a,b,c)}|w_a-w_b|^{2\Delta^c_{ab}}}
\Bigg),
\end{eqnarray}
we can view the last line of (\ref{eq:3point:variation:new}) as an effect of the re-normalization of the fields accompanied by the change of their scaling dimensions. The form of
(\ref{eq:3pt:exponential:of:T(z):explicit}) also shows that $\left|\left\langle\!\sqrt{T(z)}\right\rangle_{\hskip -3pt\hbox{\boldmath{$\scriptstyle w$}}}\right|^2_{\mathrm{reg.}}\hskip -7ptd^2z$
is invariant under translations and scaling transformations, $w_a \to w_a + t$ and $w_a \to \lambda w_a,$ although its transformation properties under the special conformal transformations are less transparent.

This nuisance is rooted in the impact of the normal ordering prescription on the transformation properties of the variation of the three-point function. 
While the correlation function containing the product $T(z_1)\ldots T(z_n)$ transform covariantly under conformal coordinate changes, this is no longer true for correlation functions containing $:\!T(z_1)\ldots T(z_n)\!:.$ It~is thus not guaranteed that the CFT deformed by the Root-$T\overline{T}$ operator, as defined in this paper, remains conformal and the obtained  formula of the variation of the three-point correlation function indicates that the classical conformal invariance of the deformed theory is indeed broken at the quantum level.

To see this more explicitly, let us notice that the normal ordering prescription (\ref{eq:normal_ordering_multiple_T}) implies that
\begin{equation}
\bra{0}:\!T^n(z)\!: = \bra{0}T^n_+(z).
\end{equation}
Since
\begin{equation}
\label{eq:Lzerominus_Tplus_commutatot}
[L_m,T^n_+(z)] = z^m\left(z\partial_z + 2n(m+1)\right)T^n_+(z), \hskip 1cm m= -1,0
\end{equation}
and thus for $L_m = L_{-1}$ or $L_0:$
\begin{equation}
\left[L_m,\mathrm{e}^{-sT_+(z)}\right]
=
\sum\limits_{n=0}^\infty \frac{(-s)^n}{n!}[L_m,T^n_+(z)]
=
z^m\left(z\frac{\partial}{\partial z} -2s \frac{\partial}{\partial s}\right)\mathrm{e}^{-sT_+(z)},
\end{equation}
we get
\begin{eqnarray}
\label{eq:Lzeroeminus_rootTplus_commutator}
\nonumber
&&
\hskip -.5cm
\left[L_m,\sqrt{T_+(z)}\right]
\; = \;
\frac{1}{\Gamma(-\alpha)}\int\limits_{0}^\infty\!ds\ s^{-\alpha-1}\,\left[L_m,\mathrm{e}^{-sT_+(z)}\right]\Big|_{\alpha \to\frac12}
\\
\nonumber
& = &
z^{m+1}\frac{\partial}{\partial z}\Big(\frac{1}{\Gamma(-\alpha)}\int\limits_{0}^\infty\!ds\ s^{-\alpha-1}\,\mathrm{e}^{-sT_+(z)}\Big|_{\alpha \to\frac12}\Big)
-
2(m+1)z^m\frac{1}{\Gamma(-\alpha)}\int\limits_{0}^\infty\!ds\, s^{-\alpha}\,\frac{\partial}{\partial s}\mathrm{e}^{-sT_+(z)}\Big|_{\alpha \to\frac12}
\\
& = &
z^m\left(z\frac{\partial}{\partial z} + m+1\right)\sqrt{T_+(z)}
\end{eqnarray}
where the last integral was computed via integrating by parts and dropping boundary terms. This shows that $\sqrt{T_+(z)}$ behaves as a dimension one operator under translation and scaling so that
we expect (and see) that insertion of the Root-$T\overline{T}$ operator in the correlation function does not spoil the translation and scaling properties of the latter. On the other hand, since
\begin{equation}
\label{eq:Lplus_Tplus_commutator}
[L_1,T_+(z)] = \left(z^2\partial_z + 4z\right)T_+(z) + 3L_{-1},
\end{equation}
and consequently
\begin{equation}
\label{eq:Lplus_multipleTplus_commutator}
\left[L_1,T^n_+(z)\right]
=
\left(z^2\partial_z + 4nz\right)T^n_+(z) -3 \sum\limits_{j=0}^{n-1}T^{n-j-1}_+(z)\partial_w T^{j}_+(z) -3nT^{n-1}_+(z)L_{-1},
\end{equation}
we see that the equality
\begin{equation}
\left[L_1,\sqrt{T_+(z)}\right] = \left(z^2\partial_z + 2z\right)\sqrt{T_+(z)}
\end{equation}
does \textbf{not} hold. This may explain the complicated behavior of (\ref{eq:3pt:exponential:of:T(z):explicit}) under special conformal transformations. 
Notice also that if the conformal invariance was preserved by the Root-$T\overline{T}$ deformation, then the formula in the next to the last line of (\ref{eq:3point:variation:new}) would give a variation of the three-point coupling constant of the original CFT model. 
A possible breakdown of the conformal symmetry by the (as defined in this work) Root-$T\overline{T}$ deformation makes such an identification doubtful.


\section{Conclusions and outlook}
\label{sec:conclusions}
\setcounter{equation}{0}
In this paper we made an attempt to define the Root-$T\bar{T}$ operator in the quantum theory. Our aim was to find a definition that would preserve the conformal invariance of the seed theory in the full quantum setting. Our rather naive approach is to first separate the holomorphic and anti-holomorphic sectors and to subsequently implement the square root by employing the Schwinger parametrization trick. In doing this, we were able to prove some interesting universal features of the perturbation. In particular, we found a shift in the scaling dimension of the operators in the correlation functions. 


Our method of regulating (in fact subtracting) short distance singularities in the $TT$ OPE  -- even if it corresponds to the standard normal ordering prescription for defining powers of the stress-energy tensor -- is not explicitly covariant under conformal transformations, which is a possible reason behind the ``anomalous'' transformation properties of the three-point function variation computed in subsection \ref{ssec:three_point_function}. On the other hand, the procedure of extracting finite term from a divergent expression (or equivalently, the choice of a normal ordering prescription) is not unique. For instance, one may define a ``corrected'' normal ordered product of two stress-tensors as
\begin{eqnarray}
\mnormalordering T^2(z)\mnormalordering\, =\; :\!T^2(z)\!: - \frac{3}{10}\partial_z^2T(z)
\end{eqnarray}
This gives
\begin{equation}
\left[L_{m},\mnormalordering T^2(z) \mnormalordering\right] = \left(z^{m+1}\partial_z+4(m+1)z^m\right)\mnormalordering T^2(w)\mnormalordering
\end{equation}
for $m= 0, \pm 1.$ It is then straightforward to explicitly check that the correlation functions
\begin{equation}
\bra{0}\mnormalordering T^2(z)\mnormalordering\, V_{\Delta}(w_2)V_{\Delta}(w_1)\ket{0}
\hskip 5mm \mathrm{and} \hskip 5mm
\bra{0}\mnormalordering T^2(z)\mnormalordering\, V_{\Delta_3}(w_3)V_{\Delta_2}(w_2)V_{\Delta_1}(w_1)\ket{0}
\end{equation}
transform covariantly under global conformal transformations. Therefore, even if the simplest and in some sense the most natural prescription fails, the original question of whether the Root-$T\overline{T}$ deformed CFT is still conformal remains to be answered, and the study of ``covariant'' normal ordering of stress-tensor products (in particular its existence and uniqueness) along with its impact on the form of the Root-$T\overline{T}$ operator is an interesting open problem to which we intend to return in the future.


Let us finally discuss the behavior of the central charge of the Virasoro algebra under the proposed Root-$T\overline{T}$ deformation.
For a CFT on a complex plane $c$  appears solely in the chiral or anti-chiral correlators of the type
\[
\bra{0}T(w_2)T(w_1)\ket{0} = \frac{c}{2}\frac{1}{(w_2-w_1)^4}.
\]
Since for any $n>0$
\[
\bra{0}:\!\overline{T}(\bar z)^n\!:T(w_2)T(w_1)\ket{0} = 0,
\]
formula (\ref{eq:root:TbarT:definition:2}) gives
\begin{equation}
\bra{0}\sqrt{\overline{T}(\bar z)}T(w_2)T(w_1)\ket{0}
=
\frac{1}{4\sqrt{\pi}}\frac{c}{(w_2-w_1)^4}\oint\limits_{\mathbb{R}_+}\!ds\ s^{-\frac32} = 0,
\end{equation}
and consequently
\begin{eqnarray}
&&
\hskip -2cm
\frac{\bra{0}\sqrt{T(z)\overline{T}(\bar z)}T(w_2)T(w_1)\ket{0}}{\bra{0}T(w_2)T(w_1)\ket{0}}
\\[2pt]
\nonumber
& = &
\frac{\bra{0}\sqrt{T(z)}T(w_2)T(w_1)\ket{0}}{\bra{0}T(w_2)T(w_1)\ket{0}}
\times
\frac{\bra{0}\sqrt{\overline{T}(\bar z)}T(w_2)T(w_1)\ket{0}}{\bra{0}T(w_2)T(w_1)\ket{0}}
=
0.
\end{eqnarray}
On the other hand, if we were to consider a conformal field theory on a Riemann surface with a metric $g$ of non-vanishing scalar curvature $R_g$, then we would need to take into account the presence of the trace anomaly,
\begin{equation}
\left\langle T^\mu_{\;\ \mu}(x)\right\rangle = \frac{c}{24\pi}R_g(x).
\end{equation}
Thus, studies of CFT on a Riemann surface of non-vanishing curvature (e.g.\ on $\mathbb{CP}^1$ rather than the complex plane)  may open up a possibility to investigate within the framework applied in this work  a flow of the conformal anomaly $c$ under the Root-$T\overline{T}$ deformation.

\subsection*{Aacknowledgments}
The authors would like to thank Ond\v{r}ej Hul\'{\i}k, Martin Ro\v{c}ek and Linus Wulff for valuable discussions. 
The work of Leszek Hadasz is supported by the Polish National Science Centre (NCN) grant nr 2023/49/B/ST2/03481. The work of Rikard von Unge is supported by the Czech science foundation GA\v{C}R through the grant``Dualities and higher derivatives" (GA23-06498S) and the grant ``Cartan supergeometries and Higher Cartan geometries" (GA24-10887S).

L.H.\ and R.vU.\ contributed equally to this work.

\appendix
\section{The semi-classical limit}
\label{ssec:semi_classical_limit}
\renewcommand{\theequation}{A.\arabic{equation}}
\setcounter{equation}{0}
Some of the analytical results obtained in the main body of the paper can be confirmed by studying the case of all scaling dimensions in the original theory being large, i.e.\ the semi-classical limit. We shall confine ourselves to the case of the three point function since we can derive the two point function as its limit. Using the differential operator introduced in (\ref{eq:derivative_exponential}) we may quite generally write
\begin{equation}
\label{eq:semithree}
\frac{\bra{0}:\!T(z)^n\!:V_3(w_3,\bar w_3)V_2(w_2,\bar w_2)V_1(w_1,\bar w_1)\ket{0}}{\bra{0}V_3(w_3,\bar w_3)V_2(w_2,\bar w_2)V_1(w_1,\bar w_1)\ket{0}}
=
\mathrm{e}^{-h}D^n\mathrm{e}^h,
\end{equation}
where
\begin{equation}
h = -\sum\limits_{a=1}^3\Delta_a\log(w_a-z) - \sum\limits_{(a,b,c)}(\Delta_a+\Delta_b-\Delta_c)\log(w_a-w_b)
\end{equation}
with $(a,b,c)\in \{(1,2,3),(2,3,1),(3,1,2)\}$ and
\[
D = \sum\limits_{a=1}^3\frac{1}{z-w_a}\frac{\partial}{\partial w_a}.
\]
Results for the expectation value of $:\!\!T(z)^n\!\!:$ in the two-point function are recovered by putting $\Delta_1=\Delta_2,\ \Delta_3=0$ and neglecting all objects referring to $w_3.$
For $\Delta_a = \Delta\cdot \delta_a$ with all $\delta_a$ of order 1 and large $\Delta$ we have
\begin{eqnarray}
&&
\hskip -1cm
\mathrm{e}^{-h}D^n\, \mathrm{e}^{h} 
=
(Dh)^n + \binom{n}{2} (Dh)^{n-2} D^2h + \binom{n}{3} (Dh)^{n-3}D^3h + 3\binom{n}{4}(Dh)^{n-4} (D^2h)^2 
\\[4pt]
\nonumber
& + &
\binom{n}{4} (Dh)^{n-4} D^4h + 10\binom{n}{5} (Dh)^{n-5} D^2h D^3 h + 15\binom{n}{6} (Dh)^{n-6}(D^2h)^3
+  \mathcal{O}\left(\Delta^{n-4}\right).
\end{eqnarray}
Since $h\propto \Delta$, terms in this expansion are organized in groups proportional to decreasing powers of $\Delta$. This is the semi-classical expansion.
With the same precision as in the equation above we have
\begin{eqnarray}
&&
\hskip -.3cm
\mathrm{e}^{-h}\mathrm{e}^{-sD}\, \mathrm{e}^{h}
=
\sum\limits_{n=0}^\infty \frac{(-s)^n}{n!}\mathrm{e}^{-h}D^n\, \mathrm{e}^{h}
\\[4pt]
\nonumber
& = &
\left(
1+ \frac{s^2}{2} D^2h - \frac{s^3}{6}D^3h + \frac{s^4}{4}\left((D^2h)^2 + \frac{1}{6}D^4h\right)
-\frac{s^5}{12}D^2h D^3h + \frac{s^6}{48}(D^2h)^3 + \ldots
\right)\mathrm{e}^{-s\hskip .5pt Dh}.
\end{eqnarray}
Therefore, up to terms of the order $\Delta^{-3}$, we can write the holomorphic part of the variation of the correlation function (\ref{eq:semithree}) as
\begin{eqnarray}
\nonumber
\label{eq:perturbative:expansion:up:to:third:order}
\frac{1}{\Gamma\left(-\alpha\right)}\int\limits_{0}^\infty\!ds\  s^{-\alpha -1} \mathrm{e}^{-h}\mathrm{e}^{-sD}\, \mathrm{e}^{h}\Big|_{\alpha = \frac12}
&=&
\sqrt{Dh}\left(1 -{\frac18\frac{D^2h}{(Dh)^2}} 
{+ \frac{1}{16}\left(\frac{D^3h}{(Dh)^3} -\frac{15}{8}\frac{(D^2h)^2}{(Dh)^4}\right)}
\right.
\\[-16pt]
\\[4pt]
\nonumber
&&
\left.
{- \frac{5}{128}\frac{D^4h}{(Dh)^4}
+\frac{35}{128}\frac{D^2hD^3h}{(Dh)^5} - \frac{315}{1024}\frac{(D^2h)^3}{(Dh)^6}}
+\ldots \right).
\end{eqnarray}

\bigskip
The semi-classical approximation allows us to compare the two possible definitions of the Root-$T\bar{T}$ operator where we either separate the holomorphic and anti-holomorphic sectors from the beginning, or we keep them together. Using the latter choice for the Root-$T\bar{T}$ operator, we would instead, to the same order as in the previous case, do the following calculation
\begin{eqnarray}
&&
\hskip -1cm
\frac{1}{n!}\mathrm{e}^{-h-\bar h}(D\bar D)^n \mathrm{e}^{h+\bar h}
=
\frac{(Dh\overline{Dh})^{n}}{n!} + \frac12\left(D^2h(\overline{Dh})^2 + (Dh)^2 \overline{D^2h}\right)\frac{(Dh\overline{Dh})^{n-2}}{(n-2)!}
\\[4pt]
\nonumber
& + &
\frac14 n(n-1) D^2h\overline{D^2h} \frac{(Dh\overline{Dh})^{n-2}}{(n-2)!}
+
\frac16\left(D^3h(\overline{Dh})^3 + \overline{D^3h}(Dh)^3\right)\frac{(Dh\overline{Dh})^{n-3}}{(n-3)!}
\\[4pt]
\nonumber
& + &
\frac14\left( (D^2h)^2 (\overline{Dh})^4 + (\overline{D^2h})^2 (Dh)^4\right)\frac{(Dh\overline{Dh})^{n-4}}{(n-4)!},
\end{eqnarray}
and consequently
\begin{eqnarray}
\label{eq:exponential:of:the:product}
&& \hskip -1cm\mathrm{e}^{-h-\bar h}\mathrm{e}^{-sD\overline{D}} \mathrm{e}^{h+\bar h}
= 
\left(1 + \frac{s^2}{2}\left(D^2h(\overline{Dh})^2 + (Dh)^2 \overline{D^2h}\right)\right)\mathrm{e}^{-s\hskip .5pt Dh\overline{Dh}}
\\[4pt]
\nonumber
& + &
\frac{s^2}{4}\left(2-4 sDh\overline{Dh} + s^2(Dh\overline{Dh})^2\right)D^2h\overline{D^2h} \mathrm{e}^{-s\hskip .5pt Dh\overline{Dh}}
\\[4pt]
\nonumber
& + &
\left(-\frac{s^3}{6}\left(D^3h(\overline{Dh})^3 + \overline{D^3h}(Dh)^3\right) + \frac{s^4}{4}\left( (D^2h)^2 (\overline{Dh})^4 + (\overline{D^2h})^2 (Dh)^4\right)\right)\mathrm{e}^{-s\hskip .5pt Dh\overline{Dh}}.
\end{eqnarray}
Explicit computation shows that
\begin{equation}
\frac{1}{\Gamma(-\alpha)}
\int\limits_{0}^\infty\!ds\ s^{-\alpha -1} \mathrm{e}^{-h-\bar h}\mathrm{e}^{-sD\overline{D}} \mathrm{e}^{h+\bar h}\Bigg|_{\alpha = \frac12}
\end{equation}
coincides (for real $\Delta$) with the modulus square of the formula (\ref{eq:perturbative:expansion:up:to:third:order}). Thus we see that it does not matter, in the semi classical approximation, whether we perform the calculations separately in the holomorphic and anti-holomorphic sectors or if we do them together. We view this equality as supporting the definition (\ref{eq:rootTbarTseparation}).

\bigskip
Notice however that the form of (\ref{eq:perturbative:expansion:up:to:third:order}) demonstrates the basic flaw of the semi-classical approach to computing the three-point function variation. The function
\begin{equation}
Dh = \prod\limits_{a=1}^3\frac{1}{z-w_a}\sum_{(a,b,c)}\frac{\Delta_a w_{ab}w_{ac}}{z-w_a},
\end{equation}
where $w_{ab} \equiv w_a-w_b,$ vanishes for $z$ satisfying the quadratic equation
\begin{equation}
\sum\limits_{(a,b,c)}\Delta_a w_{ab}w_{ac}(z-w_b)(z-w_c)=0,
\end{equation}
and the subsequent terms in the expansion (\ref{eq:perturbative:expansion:up:to:third:order}) are increasingly singular at these points. Since the exact formula for the variation of the three-point
correlation function worked out in subsection \ref{ssec:three_point_function} does not contain this singularity, its presence in (\ref{eq:perturbative:expansion:up:to:third:order}) should be interpreted as a breakdown of the
semi-classical expansion.

On the other hand, for the two-point function we have
\begin{equation}
Dh = \frac{(w_2-w_1)^2}{(z-w_2)^2(z-w_1)^2}\,\Delta,
\end{equation}
so the perturbation series breaks down only for $z\to\infty.$ It can be re-summed with the result:
\begin{equation}
\frac{\bra{0}\sqrt{T(z)}V_\Delta(w_2,\bar w_2)V_\Delta(w_1,\bar w_1)\ket{0}}{\bra{0}V_\Delta(w_2,\bar w_2)V_\Delta(w_1,\bar w_1)\ket{0}}
=
\frac{\sqrt{2}\,\Gamma\left(\frac{\Delta+1}{2}\right)}{\Gamma\left(\frac{\Delta}{2}\right)}\left(\frac{1}{\sqrt{(z-w_2)^2}}+\frac{1}{\sqrt{(z-w_1)^2}}\right) + \mathrm{reg.},
\end{equation}
where `reg.'\ stands for terms finite for $z \in \mathbb{C}.$ This result is in accord with the non-perturbative computation presented in subsection \ref{ssec:two_point_function}. One can also prove that
\begin{equation}
\label{eq:T_to_n:two_point:w2_to_infinity}
\lim\limits_{w_2\to \infty}\frac{\bra{0} :\!T(z)^n\!: V_{\Delta}(w_2,\bar w_2)V_{\Delta}(w_1,\bar w_1)\ket{0}}{\bra{0} V_{\Delta}(w_2,\bar w_2)V_{\Delta}(w_1,\bar w_1)\ket{0}}
=
\frac{2^n\Gamma\left(\frac{\Delta}{2}+n\right)}{\Gamma\left(\frac{\Delta}{2}\right)(z-w_1)^{2n}},
\end{equation}
and consequently that
\begin{eqnarray}
\nonumber
\lim\limits_{w_2\to \infty}\frac{\bra{0}:\!\mathrm{e}^{-sT(z)}\!: V_{\Delta}(w_2,\bar w_2)V_{\Delta}(w_1,\bar w_1)\ket{0}}{\bra{0} V_{\Delta}(w_2,\bar w_2)V_{\Delta}(w_1,\bar w_1)\ket{0}}
& = &
\sum\limits_{n=0}^\infty \frac{\Gamma\left(\frac{\Delta}{2}+n\right)}{n!\Gamma\left(\frac{\Delta}{2}\right)}\left(-\frac{2s}{(z-w_1)^2}\right)^n
\\
& = &
\left(1+\frac{2s}{(z-w_1)^2}\right)^{-\frac{\Delta}{2}}.
\end{eqnarray}
These relations can now be used to further illustrate the convenience of our choice of definition of the Root-$T\bar{T}$ operator.
The identity
\begin{equation}
\lim\limits_{w_2,\bar w_2\to \infty}\frac{\bra{0} :\!T(z)^n\!: :\!\overline{T}(\bar z)^n\!:V_{\Delta}(w_2,\bar w_2)V_{\Delta}(w_1,\bar w_1)\ket{0}}{\bra{0} V_{\Delta}(w_2,\bar w_2)V_{\Delta}(w_1,\bar w_1)\ket{0}}
=
\frac{4^n\Gamma^2\left(\frac{\Delta}{2}+n\right)}{\Gamma^2\left(\frac{\Delta}{2}\right)|z-w_1|^{4n}},
\end{equation}
a direct consequence of (\ref{eq:T_to_n:two_point:w2_to_infinity}), shows that the radius of convergence of the series
\begin{equation}
\sum\limits_{n=0}^{\infty}\frac{(-s)^n}{n!}\lim\limits_{w_2,\bar w_2\to \infty}\frac{\bra{0} :\!T(z)^n\!: :\!\overline{T}(\bar z)^n\!:V_{\Delta}(w_2,\bar w_2)V_{\Delta}(w_1,\bar w_1)\ket{0}}{\bra{0} V_{\Delta}(w_2,\bar w_2)V_{\Delta}(w_1,\bar w_1)\ket{0}}
\end{equation}
is equal to zero. This implies that while the state $V_{\Delta}(w_2,\bar w_2)V_{\Delta}(w_1,\bar w_1)\ket{0}$ is in the domain of the operator $:\!\mathrm{e}^{-sT(z)}\!:$ this is no longer true for the operator
$:\!\mathrm{e}^{-sT(z)\overline{T}(\bar z)}\!:$, which gives additional motivation for our choice of separating the holomophic and anti holomorphic sectors in the definition of the Root-$T\overline{T}$ operator.

\section{Free scalar CFT}
\label{ssec:free_scalar}
\renewcommand{\theequation}{B.\arabic{equation}}
\setcounter{equation}{0}
A free, massless, scalar field is a model singled out by the fact that -- at the classical level -- its Root-$T\overline T$ deformation is trivial and reduces to the re-scaling of the classical action. In effect, because of the simplicity of the model, we can independently check if the large scaling dimensions limit of the general formulas worked out in the main body of the paper coincide with the corresponding formulas worked out from the classically deformed action functional. 

If we consider this theory on the Minkowski cylinder with circumference $2\pi,$ then its classical action reads
\begin{equation}
\label{eq:free:field:action}
S_g[\phi] = \frac{g}{2}\int\!dt\int\limits_0^{2\pi}\!d\sigma\ \left((\partial_t\phi)^2 - (\partial_\sigma\phi)^2\right),
\end{equation}
where we have explicitly introduced the normalization constant $g.$ 

When we analytically continue $t \to i\tau$ and pass to the complex coordinates
\begin{equation}
\mathrm{e}^{-i(t-\sigma)}\ \to \ \mathrm{e}^{\tau + i\sigma} \equiv \mathrm{e}^{\tau^+} = z,
\hskip 1cm
\bar z =  \mathrm{e}^{-i(t+\sigma)} \to \mathrm{e}^{\tau -i\sigma} \equiv \mathrm{e}^{\tau^-} = \bar z,
\end{equation}
then then the general solution of the classical equation of motion can be written as
\begin{equation}
\partial\phi(z) = - \frac{1}{\sqrt{2\pi g}}\sum\limits_{n\in\mathbb Z}\frac{a_n}{z^{n+1}}.
\end{equation}
The holomorphic component of the energy-momentum tensor computed from the action (\ref{eq:free:field:action}) reads
\begin{equation}
\label{eq:Tclassical}
T(z) = 2\pi g (\partial\phi(z))^2,
\end{equation}
and consequently
\begin{equation}
\epsilon\int\! d^2z\ \sqrt{T(z)\overline{T}(\bar z)}  
=
\frac{\epsilon\pi g}{2}\int\! d\tau\wedge d\sigma \left((\partial_\tau \phi(\tau,\sigma))^2 + (\partial_\sigma \phi(\tau,\sigma))^2\right)
\equiv 
\epsilon\pi S_g^{\mathrm{\scriptscriptstyle{E}}}[\phi],
\end{equation}
where the superscript E denotes the Euclidean action. The classical effect of the Root-$T\overline{T}$ deformation thus reduces to the scaling of the action's normalization constant $g \to g' = (1+\epsilon\pi)g.$

When we quantize the theory defined by the action (\ref{eq:free:field:action}) we arrive at the OPE
\begin{equation}
\partial\phi(z)\partial\phi(w) \sim  \frac{1}{4\pi g}\frac{1}{(z-w)^2},
\end{equation}
and the quantum counterpart of (\ref{eq:Tclassical}) is 
\begin{equation}
\label{eq:free:scalar:holomorphic:T}
T(z) = 2\pi g\lim\limits_{w\to z}\left(\partial\phi(w)\partial\phi(z)-\frac{1}{4\pi g}\frac{1}{(w-z)^2}\right).
\end{equation}
It is immediate to check that $T(z)$ is correctly normalized and the (Virasoro) algebra of its modes has central charge $c=1,$
\begin{equation}
T(z)T(w) \sim \frac12 \frac{1}{(z-w)^4} + \frac{2T(w)}{(z-w)^2}+ \frac{\partial T(w)}{z-w}.
\end{equation}
Next, for the normal ordered exponential, defined as
\begin{equation}
\label{eq:normal:ordered:exponential}
\mathsf{E}^\alpha(w,\bar w)
=
\mathrm{e}^{\alpha\phi_<(w,\bar w)}\,\mathrm{e}^{\alpha\phi_>(w,\bar w)}\,
\mathrm{e}^{\frac{\alpha}{\sqrt{2\pi g}}q}\,|z|^{-\frac{2i\alpha}{\sqrt{2\pi g}} p},
\end{equation}
where $\phi_<(w,\bar w)$ (resp.\ $\phi_>(w,\bar w)$) contain solely creation (resp.\ annihilation) operators while $q$ and $p=-ia_0$ are zero modes,
we obtain
\begin{equation}
T(z)\mathsf{E}^\alpha(w,\bar w) \sim  \left(\frac{\alpha^2}{8\pi g}\frac{1}{(z-w)^2} + \frac{1}{z-w}\frac{\partial}{\partial w}\right)\mathsf{E}^\alpha(w,\bar w).
\end{equation}
This formula shows that $\mathsf{E}^\alpha(w,\bar w)$ is a primary field with conformal dimension 
\begin{equation}
\label{eq:free:field:scaling:dimension}
\Delta_\alpha = \frac{\alpha^2}{8\pi g}
\end{equation}
and therefore, for $g\to (1+\epsilon\pi)g,$ we get 
\begin{equation}
\delta_\epsilon \Delta_\alpha = \frac{\alpha^2}{8\pi g}\left(\frac{1}{1+\epsilon\pi} -1\right) = -\epsilon\pi \Delta_\alpha
+ \mathcal{O}\left(\epsilon^2\right),
\end{equation}
in agreement with the large $\Delta$ limit of (\ref{eq:scaling:dimension:variation}).

Notice also that the zero mode of the canonical momentum $\pi(t,\sigma) = g\dot\phi(t,\sigma)$ is equal to
\begin{equation}
\pi_0 = \sqrt{\frac{2g}{\pi}}\,p
\end{equation}
and computing the commutator
\begin{equation}
\left[\pi_0,\mathsf{E}^\alpha(w,\bar w) \right] = - \frac{i\alpha}{2\pi}\mathsf{E}^\alpha(w,\bar w)
\end{equation}
we conclude that the momentum related to the operator $\mathsf{E}^\alpha(w,\bar w)$ is independent of the normalization constant $g.$

Now, in the semi-classical limit of all $\Delta_a$ being large the three-point function variation 
$\left|\left\langle\!\sqrt{T(z)}\right\rangle_{\hskip -3pt\hbox{\boldmath{$\scriptstyle w$}}}\right|^2_{\mathrm{reg.}}$  given by eq.\ (\ref{eq:3pt:variation:regularized}) tends to
\begin{equation}
\label{eq:3pt_classical_limit}
\left|\sum\limits_{a=1}^3\frac{\Delta_a}{(z-w_a)^2}- \sum\limits_{(a,b,c)}\frac{\Delta^c_{ab}}{(z-w_a)(z-w_b)}\right|
-
\frac12\sum\limits_{(a,b,c)}\frac{\Delta^c_{ab}|w_a-w_b|^2}{|z-w_a|^2|z-w_b|^2}.
\end{equation}
In the free field theory the momentum conservation implies that in the three-point function $\sqrt{\Delta_3} =  \sqrt{\Delta_1}\pm\sqrt{\Delta}_2.$
Then
\begin{equation}
\left|\sum\limits_{a=1}^3\frac{\Delta_a}{(z-w_a)^2}- \sum\limits_{(a,b,c)}\frac{\Delta^c_{ab}}{(z-w_a)(z-w_b)}\right|
=
\left|\frac{\sqrt{\Delta_1}}{z-w_1} \pm \frac{\sqrt{\Delta_2}}{z-w_2} - \frac{\sqrt{\Delta_1}\pm\sqrt{\Delta_2}}{z-w_3}\right|^2
\end{equation}
and applying the identity
\begin{eqnarray}
&&
\hskip -.5cm
\left|\frac{a}{z-w_1}+ \frac{b}{z-w_2} - \frac{a+b}{z-w_3}\right|^2
\\[2pt]
\nonumber
& = & 
a(a+b)\left|\frac{w_{13}}{(z-w_1)(z-w_3)}\right|^2
+
b(a+b)\left| \frac{w_{23}}{(z-w_2)(z-w_3)}\right|^2
-
ab\left|\frac{w_{12}}{(z-w_1)(z-w_2)}\right|^2,
\end{eqnarray}
valid for real $a$ and $b,$ one immediately shows that in such a case
(\ref{eq:3pt_classical_limit}) 
vanishes identically.  This is expected: the three-point coupling constant in the free boson CFT has the form of a Dirac delta enforcing the conservation of the momentum and, since the re-scaling of the coefficient in front of the action does not affect the momenta associated with primary fields, this relation stays intact.

Let us close this Appendix with a remark on the equivalence (for the free, massless boson) of  scaling the coefficient in front of the action versus the Root-$T\overline{T}$ deformation. In the former case the change of the primary field's scaling dimension $\Delta$ is strictly proportional to $\Delta$ itself. On the other hand, the  Root-$T\overline{T}$ deformation as defined in our work -- both for the standard and ``improved'' normal ordering procedures -- results in corrections to the formula $\delta\Delta \propto \Delta$  which are negligible only for $\Delta \gg 1$ (cf Eq.\ (\ref{eq:scaling:dimension:variation})). This shows that deforming the theory with the Root-$T\overline{T}$ as defined in this paper gives a different result from just naively rescaling the coefficient in front of the action. The question of what model one obtains through the Root-$T\overline{T}$ deformation of the free scalar theory remains a non-trivial interesting, open problem.

\section{Proof of properties of the $\sqrt{T}$ variation of the two-point function}
\label{app:analytic_properties}
\renewcommand{\theequation}{C.\arabic{equation}}
\setcounter{equation}{0}
The integral
\begin{equation}
\label{eq:diveregent:two:point}
I_\alpha(x_2,x_1) = 
\frac{1}{2\Gamma(-\alpha)\left(\mathrm{e}^{2\pi i\alpha}-1\right)}\oint\limits_{\mathbb{R}_+}\!ds\ s^{-\alpha-1}\prod\limits_{j=1}^2\Bigg(\frac{x_j}{\sqrt{x_j^2+2s}}\Bigg)^\Delta\left(\frac{x_2-x_1}{\sqrt{x_2^2+2s}-\sqrt{x_1^2+2s}}\right)^{2\Delta}\,
\end{equation}
with the integration contour encircling the positive real semi-axis in the counterclockwise direction, is convergent for $\Re\,\alpha < 0.$ For $\alpha = \frac12$ we can rewrite it as
\begin{eqnarray}
\nonumber
&&
\hskip -.5cm
I(x_2,x_1) 
=
-\frac{1}{2\sqrt{\pi}}\lim\limits_{\epsilon\to 0}
\Bigg(
\int\limits_\epsilon^\infty\!ds\ s^{-3/2}\prod\limits_{j=1}^2\Bigg(\frac{x_j}{\sqrt{x_j^2+2s}}\Bigg)^\Delta\left(\frac{x_2-x_1}{\sqrt{x_2^2+2s}-\sqrt{x_1^2+2s}}\right)^{2\Delta}
-
\frac{2}{\sqrt{\epsilon}}
\Bigg)
\\
& = &
-\frac{1}{2\sqrt{\pi}}\lim\limits_{\epsilon\to 0}
\Bigg(
\int\limits_\epsilon^\infty\!ds\ s^{-3/2}\left(\frac{x_1x_2}{(x_2+x_1)^2}\frac{\big(\sqrt{x_2^2+2s}+\sqrt{x_1^2+2s}\big)^2}{\sqrt{x_1^2+2s}\sqrt{x_1^2+2s}}\right)^{\Delta}
-
\frac{2}{\sqrt{\epsilon}}
\Bigg).
\end{eqnarray}

Notice now that for $x_i^2, s \in \mathbb{R}_+$ the function
\begin{equation}
g(s) = \frac{\left(\sqrt{x_1^2+2s}+\sqrt{x_2^2+2s}\right)^2}{\sqrt{x_1^2+2s}\sqrt{x_2^2+2s}}
\end{equation}
is positive and monotonically decreasing,
\begin{equation}
\frac{dg(s)}{ds} = - \frac{\left(x_2^2-x_1^2\right)^2}{\left(\sqrt{x_1^2+2s}\sqrt{x_2^2+2s}\right)^3},
\end{equation}
and thus, for $\Delta > 0:$
\begin{eqnarray*}
&&
\hskip -2cm
\left(\frac{x_1x_2}{(x_1+x_2)^2}\right)^{\Delta}\int\limits_{x_2}^\infty\!ds\ s^{-3/2}\Big(g(s)\Big)^\Delta
\leq
\left(\frac{x_1x_2}{(x_1+x_2)^2}g(x_2)\right)^{\Delta}\int\limits_{x_2}^\infty\!ds\ s^{-3/2}
\\
& = &
\frac{2}{\sqrt{x_2}}\left(\frac{x_1x_2\left(\sqrt{x_1^2+2x_2}+\sqrt{x_2^2+2x_2}\right)^2}{(x_1+x_2)^2\sqrt{x_1^2+2x_2}\sqrt{x_2^2+2x_2}}\right)^{\Delta}
\stackrel{x_2 \gg x_1}{\sim}
\frac{2}{\sqrt{x_2}}\left(\frac{x_1}{\sqrt{2x_2}}\right)^\Delta\hskip -4pt.
\end{eqnarray*}
Therefore
\begin{equation}
\lim\limits_{x_2\to\infty}
\int\limits_{x_2}^\infty\!ds\ s^{-3/2}\left(\frac{x_1x_2}{(x_2+x_1)^2}\frac{\big(\sqrt{x_2^2+2s}+\sqrt{x_1^2+2s}\big)^2}{\sqrt{x_1^2+2s}\sqrt{x_1^2+2s}}\right)^{\Delta}
= 0
\end{equation}
and, since for $\epsilon \leq s \leq x_2:$ 
\[
\lim\limits_{x_2\to\infty}\Bigg(\frac{x_2}{\sqrt{x_2^2+2s}}\Bigg)^\Delta\left(\frac{\sqrt{x_2^2+2s}+\sqrt{x_1^2+2s}}{x_2+x_1}\right)^{2\Delta} = 1,
\]
we get
\begin{eqnarray}
&&\hskip -10mm
\lim\limits_{x_2\to\infty}I(x_2,x_1)
=
-\frac{1}{2\sqrt{\pi}}\lim\limits_{\epsilon\to 0}
\Bigg(
x_1^\Delta\int\limits_\epsilon^\infty\! \frac{s^{-3/2}\,ds}{\left(x_1^2+2s\right)^{\Delta/2}}
-
\frac{2}{\sqrt{\epsilon}}
\Bigg)
=
\frac{x_1^\Delta}{4\sqrt{\pi}}\oint\limits_{\mathbb R_+}\! \frac{s^{-3/2}\,ds}{\left(x_1^2+2s\right)^{\Delta/2}}
\\
\nonumber
& = & 
\frac{x_1^\Delta}{4\sqrt{\pi}}\oint\limits_{\mathbb R_+}\! \frac{s^{-\alpha-1}\,ds}{\left(x_1^2+2s\right)^{\Delta/2}}\Bigg|_{\alpha = \frac12}
=
-\frac{x_1^\Delta}{2\sqrt{\pi}}\int\limits_0^{\infty}\! \frac{s^{-\alpha-1}\,ds}{\left(x_1^2+2s\right)^{\Delta/2}}\Bigg|_{\alpha = \frac12}
=
\frac{\sqrt{2}\,\Gamma\left(\frac{\Delta+1}{2}\right)}{\Gamma\left(\frac{\Delta}{2}\right)}\frac{1}{\sqrt{(w_1-z)^2}},
\end{eqnarray}
where, in evaluating the integral, we first expressed it as a contour integral, then we presented the integrand as a analytic function of $\alpha,$ 
deformed (for $\Re\alpha < 0$) the integration contour onto the real semi-axis, evaluated the integral for $\Re\alpha < 0$ and computed the resulting formula for $\alpha = \frac12.$ The result proves (\ref{eq:w2_to_infinity})

In order to prove (\ref{eq:z_to_w1}) we first take $s \leq x_1$ and $x_1 \ll x_2 \sim 1.$ In such a situation:
\begin{equation}
\Bigg(\frac{x_2}{\sqrt{x_2^2+2s}}\Bigg)^\Delta\left(\frac{\sqrt{x_2^2+2s}+\sqrt{x_1^2+2s}}{x_2+x_1}\right)^{2\Delta} \approx 1
\end{equation}
and
\begin{eqnarray}
&&
\hskip -.5cm
\frac{-1}{2\sqrt{\pi}}\lim\limits_{\epsilon\to 0}
\Bigg(
\int\limits_\epsilon^{x_1}\!ds\ s^{-3/2}\prod\limits_{j=1}^2\Bigg(\frac{x_j}{\sqrt{x_j^2+2s}}\Bigg)^\Delta\left(\frac{x_2-x_1}{\sqrt{x_2^2+2s}-\sqrt{x_1^2+2s}}\right)^{2\Delta}
-
\frac{2}{\sqrt{\epsilon}}
\Bigg)
\\
& \approx &
\nonumber
\frac{-1}{2\sqrt{\pi}}\lim\limits_{\epsilon\to 0}
\Bigg(
\int\limits_\epsilon^{x_1}\!ds\ s^{-3/2}\Bigg(\frac{x_1}{\sqrt{x_1^2+2s}}\Bigg)^\Delta
-
\frac{2}{\sqrt{\epsilon}}
\Bigg)
=
\frac{-1}{\sqrt{2\pi}}\lim\limits_{\epsilon\to 0}
\Bigg(
\frac{1}{x_1}\int\limits_{\frac{2\epsilon}{x_1^2}}^{\frac{2}{x_1}}\!dt\ \frac{t^{-3/2}}{(1+t)^{\frac{\Delta}{2}}}
-
\sqrt{\frac{2}{\epsilon}}
\Bigg)
\\
& = &
\frac{-1}{2\sqrt{2\pi}}\frac{1}{x_1} \int\limits_{C}\!dt\ \frac{t^{-3/2}}{(1+t)^{\frac{\Delta}{2}}}
\end{eqnarray}
where the integration contour starts at $t =\frac{2}{x_1}\,{\mathrm e}^{2\pi i},$ encircles in the clockwise direction the branch point at $t=0$ and ends at $t = \frac{2}{x_1}.$
Analytically continuing $t^{-\frac32}\to t^{-\alpha-1},$ computing the resulting integral and continuing back to $\alpha = \frac12$ we get in the limit $x_1\to 0:$
\begin{equation}
\frac{-1}{2\sqrt{2\pi}}\int\limits_{C}\!dt\ \frac{t^{-3/2}}{(1+t)^{\frac{\Delta}{2}}} = \frac{\sqrt{2}\,\Gamma\left(\frac{\Delta+1}{2}\right)}{\Gamma\left(\frac{\Delta}{2}\right)} + 
\mathcal{O}(x_1).
\end{equation}
Now, for any $s \geq x_1:$
\begin{equation}
\frac{x_1x_2}{(x_2+x_1)^2}\frac{\big(\sqrt{x_2^2+2s}+\sqrt{x_1^2+2s}\big)^2}{\sqrt{x_1^2+2s}\sqrt{x_1^2+2s}}
\leq
\frac{x_1x_2}{(x_2+x_1)^2}\frac{\big(\sqrt{x_2^2+2x_1}+\sqrt{x_1^2+2x_1}\big)^2}{\sqrt{x_1^2+2x_1}\sqrt{x_1^2+2x_1}}
\leq 1
\end{equation}
and consequently, for $\Delta > 0:$
\begin{eqnarray}
&&
\hskip -2cm
\lim\limits_{x_1\to 0}
\Bigg|-\frac{x_1}{2\sqrt{\pi}}
\int\limits_{x_1}^\infty\!ds\ s^{-3/2}\left(\frac{x_1x_2}{(x_2+x_1)^2}\frac{\big(\sqrt{x_2^2+2s}+\sqrt{x_1^2+2s}\big)^2}{\sqrt{x_1^2+2s}\sqrt{x_1^2+2s}}\right)^{\Delta}
\Bigg|
\\
\nonumber
& \leq &
\lim\limits_{x_1\to 0}
\Bigg|-\frac{x_1}{2\sqrt{\pi}}
\int\limits_{x_1}^\infty\!ds\ s^{-3/2}
\Bigg|
=
\lim\limits_{x_1\to 0}\Bigg|-\sqrt{\frac{x_1}{\pi}}\Bigg| = 0,
\end{eqnarray}
and we conclude that for $x_1\to 0:$
\begin{equation}
x_1 I(x_2,x_1) = \frac{\sqrt{2}\,\Gamma\left(\frac{\Delta+1}{2}\right)}{\Gamma\left(\frac{\Delta}{2}\right)} + o(1).
\end{equation}
Let us finally consider the limit of $-z \equiv \varepsilon^{-1} \gg 1.$ We have:
\begin{eqnarray}
&&
\hskip -.4cm 
I_{-\frac12}(w_2+\epsilon^{-1},w_1+\varepsilon^{-1})
\\
\nonumber
& = &
-\frac{\varepsilon}{2\sqrt{2\pi}}\oint\limits_{\mathbb{R}_+}\hskip -2pt dt\,
t^{-3/2}\left(\frac{(1+\varepsilon w_1)(1+\varepsilon w_2)}{(2+\varepsilon w_1+\varepsilon w_2)^2}\frac{\big(\sqrt{(1+\varepsilon w_1)^2+t}+\sqrt{(1+\varepsilon w_2)^2+t}\big)^2}{\sqrt{(1+\varepsilon w_1)^2+t}\sqrt{(1+\varepsilon w_2)^2+t}}\right)^{\Delta}.
\end{eqnarray}
For $\varepsilon \to 0:$
\begin{equation}
\frac{(1+\varepsilon w_1)(1+\varepsilon w_2)}{(2+\varepsilon w_1+\varepsilon w_2)^2}\frac{\big(\sqrt{(1+\varepsilon w_1)^2+t}+\sqrt{(1+\varepsilon w_2)^2+t}\big)^2}{\sqrt{(1+\varepsilon w_1)^2+t}\sqrt{(1+\varepsilon w_2)^2+t}}
=
1+\mathcal{O}\left(\epsilon^2\right)
\end{equation}
and therefore
\begin{equation}
\lim\limits_{\varepsilon\to 0}
\oint\limits_{\mathbb{R}_+}\hskip -2pt dt\,
t^{-3/2}
\Bigg\{
\frac{(1+\varepsilon w_1)(1+\varepsilon w_2)}{(2+\varepsilon w_1+\varepsilon w_2)^2}\frac{\big(\sqrt{(1+\varepsilon w_1)^2+t}+\sqrt{(1+\varepsilon w_2)^2+t}\big)^2}{\sqrt{(1+\varepsilon w_1)^2+t}\sqrt{(1+\varepsilon w_2)^2+t}}
-
\mathrm{e}^{-\varepsilon^2 t}
\Bigg\}
=
0.
\end{equation}
Moreover
\begin{equation}
-\frac{\varepsilon}{4\sqrt{\pi}}\oint\limits_{\mathbb{R}_+}\hskip -2pt dt\  t^{-\frac32}\mathrm{e}^{-\varepsilon^2 t} = \frac12\varepsilon^2
\end{equation}
so that for $\varepsilon \to 0:$
\begin{equation}
I_{-\frac12}(w_2+\epsilon^{-1},w_1+\varepsilon^{-1}) = o\left(\varepsilon\right).
\end{equation}
This proves (\ref{eq:z_to_infinity}).

\end{document}